# Temporal genomics help in deciphering neutral and adaptive patterns in the contemporary evolution of kelp populations

Lauric Reynes[1,2*], Louise Fouqueau[1,3], Didier Aurelle[4,5], Stéphane Mauger[1] Christophe Destombe[1], Myriam Valero[1]
Corresponding author: lreynes@hawaii.edu

[1]IRL 3614, CNRS, Sorbonne Université, Pontificia Universidad Católica de Chile, Universidad Austral de Chile, Station Biologique de Roscoff, Roscoff 29688, France
[2] Present address: The Sherwood Laboratory, School of Life Sciences, University of Hawai'i at Mānoa Honolulu, Hawai'i 96822, U.S.A
[3] Present address: Institute of Science and Technology Austria, 3400 Klosterneuburg, Austria
[4]Aix-Marseille Université, Université de Toulon, CNRS, IRD, MIO, 13288 Marseille, France
[5]Institut de Systématique Évolution Biodiversité (ISYEB, UMR 7205), Muséum National d'Histoire Naturelle, CNRS, EPHE, Sorbonne Université, Paris, France

## ABSTRACT

The impact of climate change on populations will be contingent upon their contemporary adaptive evolution. In this study, we investigated the contemporary evolution of four populations of the cold-water kelp *Laminaria digitata* by analysing their spatial and temporal genomic variation using ddRAD-sequencing. These populations were sampled from the center to the southern margin of its north-eastern Atlantic distribution at two-time points, spanning at least two generations. Through genome scans for local adaptation at a single time point, we identified candidate loci that showed clinal variation correlated with changes in sea surface temperature (SST) along latitudinal gradients. This finding suggests that SST may drive the adaptive response of these kelp populations, although factors such as species' demographic history should also be considered. Additionally, we performed a simulation approach to distinguish the effect of selection from genetic drift in allele frequency changes over time. This enabled the detection of loci in the southernmost population that exhibited temporal differentiation beyond what would be expected from genetic drift alone: these are candidate loci which could have evolved under selection over time. In contrast, we did not detect any outlier locus based on temporal differentiation in the population from the North Sea, which also displayed low and decreasing levels of genetic diversity. The diverse evolutionary scenarios observed among populations can be attributed to variations in the prevalence of selection relative to genetic drift across different environments. Therefore, our study highlights the potential of temporal genomics to offer valuable insights into the contemporary evolution of marine foundation species facing climate change.

# 1. INTRODUCTION

Genetic variation results from the interplay of selection, mutation, migration and genetic drift, and is essential to assess the ability of natural populations to cope with rapid climate change (Dawson et al., 2011; Hoban et al., 2020). Amongst these forces, genetic drift is expected to decrease within-population genetic variation, while the between-population genetic variation may increase when different alleles fixate across populations. Nonetheless, strong drift is generally considered to hinder the implementation of local adaptation (Le Corre & Kremer, 2012), yet local adaptation remains an important source of intraspecific genetic variation when selection pressures vary across heterogeneous environment and species range. This is also expected in the presence of a substantial asymmetry in gene flow, as it could result in a situation where the local population's genetic makeup is overwhelmed ('swamping', Lenormand, 2002, but see Durkee et al., this issue; Sexton et al., this issue). Nonetheless, gene flow could also facilitate adaptation by increasing local genetic diversity and by decreasing the relative effect of genetic drift through the increase in effective population size (Ne).

Large Ne is crucial for safeguarding genetic diversity (Bürger and Lynch, 1995; Frankham, 1995). Yet with climate change and related extreme events, populations may encounter more frequent reductions and fluctuations in Ne over time (Holt, 1990; Pauls et al., 2013). Any decline in Ne can compromise the effectiveness of selection compared to genetic drift (see Charlesworth, 2009 for a review), thus increasing the likelihood of stochastically losing alleles, which can be problematic, especially for those associated with fitness-related traits. Estimation of Ne through temporal methods across contemporary time frames (referred to as 'contemporary Ne', Waples, 1989) offers a current assessment of the relative influence of selective compared to neutral processes (Hare et al., 2011). Amongst the methods using temporal data, the one based on allele frequency changes has been shown to give more accurate inferences of contemporary Ne and to enhance the power to detect its declines particularly when historical population sizes were large (e.g., Nadachowska-Brzyska et al., 2021; Reid & Pinsky, 2022). Nonetheless, these methods usually neglect the effect that other evolutionary processes have on allelic fluctuations (Jorde & Ryman, 2007). While these methods can be beneficial for assessing Ne in spatially separated populations with minimal gene flow, their efficiency may decrease when local gene flow occurs. Local gene flow from populations, even those not sampled, has the potential to shape variation in allele frequency, implying that local Ne estimate may be confounded with gene flow. In this scenario, a joint estimate of Ne and gene flow should be considered instead (e.g., Wang & Whitlock, 2003).

Genome-wide screening of genetic variation supports that populations can adapt to local environmental conditions despite substantial levels of genetic drift. Indeed, candidate loci for positive selection have been identified in relatively small and structured populations (e.g., Funk et al., 2016; Perrier et al., 2017; Leal et al., 2021; Pratlong et al., 2021), raising questions regarding the relative contribution of selection and drift in such small populations. Additionally, genome scan approaches, such as those using $F_{ST}$, can be biased by high variance in $F_{ST}$ amongst loci as well as by processes other than selection leading to spatial variation in allele frequencies (Storz, 2005; Bierne et al., 2013; Lotterhos & Whitlock, 2015; Hoban et al., 2016). Apart from a spatial approach which comprises samples from populations distributed across the space, selection can also be studied at the temporal scale, by comparing data from different time points of a given population. Temporal genomics allows distinguishing signatures of selection from neutral processes through the analysis of genomic variation over time (see Clark et al., 2023 for a review). This framework offers valuable insights into the species' adaptive response, by assessing adaptive patterns (e.g., Therkildsen et al., 2013a, b; Frachon et al., 2017) and facilitating the monitoring of genetic erosion experienced by populations in response to rapidly changing environments (Jensen & Leigh, 2022). The underlying premise of temporal genomics is that alleles under selection exhibit a consistent directional trend in frequency variation, unlike the random changes expected with genetic drift. Hence comparing temporal genetic differentiation expected by genetic drift using simulations (Goldringer & Bataillon, 2004) or assessing temporal covariance in allele frequencies (Buffalo & Coop, 2019, 2020), enables to explore the presence of any response to contemporary selection. While the method of Buffalo & Coop (Buffalo & Coop, 2019, 2020) appears more suitable for polygenic adaptation (very large number of loci with each having a small effect on the trait), $F_{ST}$ methods are better suited when the trait under interest involves few loci under strong selection.

The laminariales, commonly known as kelps, represent an interesting model for investigating the interplay between selection and genetic drift in the contemporary evolution of populations facing rapid climate warming. Climate warming has been identified as one of the leading causes behind population declines, local extinctions and range shifts in kelp species (Araújo et al., 2016; Arafeh-Dalmau et al., 2019; Filbee-Dexter et al., 2020). Evaluating the vulnerability of these cold-temperate water species to local extinction requires a comprehensive understanding of their capacity to adapt to warmer conditions. Recent studies on North Atlantic populations of *Laminaria digitata* have suggested that range edges populations might exhibit lower sensitivity to heat stress compared to central ones (Liesner et al., 2020; Schimpf et al., 2022). This is particularly true of the populations of Quiberon (South Brittany), located at the species' southern margin, and Helgoland (North Sea), a highly isolated population situated at the edge of the South-Atlantic cluster (Liesner et al., 2020), where seawater warming has intensified over the past few years (Pehlke & Bartsch, 2008). Although significant progress has been made in understanding the genetic basis of heat tolerance in kelp species (Guzinski et al., 2020; Mao et al., 2020; Vranken et al., 2021), identifying candidate loci for local adaptation by genome scans from a spatial approach remains challenging due to high levels of neutral differentiation in these species. Indeed, kelp species usually display considerable genetic structure attributed to limited dispersal capacities and seascape features (Coleman et al., 2011; Brennan et al., 2014; Assis et al., 2022; Fouqueau et al., 2023). This genetic structure can also be exacerbated in the presence of isolated and marginal populations mainly characterized by low to moderate effective population sizes (Fouqueau, 2021). Based on the current state of knowledge, the relative contributions of local adaptation and phenotypic plasticity to the observed thermal tolerance are still uncertain for most kelp species.

The primary objective of our study is to assess the relative contributions of local selection and genetic drift in the contemporary evolution of the kelp *Laminaria digitata* across both spatial and temporal dimensions. To achieve this, we used a reduced representation sequencing technique (double digest Restriction site Associated DNA sequencing, or ddRAD-seq, Peterson et al., 2012) on four populations, with three of these populations sampled at two-time points. Our objectives encompassed the following aims: i) Assessing whether temporal genomics on contemporary scales can be used to evaluate the relative contributions of genetic drift and selection within populations, ii) Testing the hypothesis that the high rate of genetic drift which is expected in small and isolated populations (Helgoland and Quiberon) may impede the efficiency of selection in these populations, iii) Conducting genome scans for detecting footprints of adaptation to local thermal conditions. Overall, our study introduces a novel framework, combining spatial and temporal genomics to gain deeper insights into the contemporary evolution of this fundamental marine species in response to rapid environmental changes.

## 2. MATERIAL AND METHODS

### 2.1    Model species and sampling

Similar to all laminarian kelps, *Laminaria digitata* follows a dioecious haplo-diplontic life cycle, consisting of an alternation between microscopic haploid gametophytes (males or females) and macroscopic diploid sporophytes. Sex determination takes place during meiosis, through a UV sexual system (Coelho et al., 2018), while the diploid sporophyte remains asexual. The lifespan of sporophytes was reported to be four to six years (Bartsch et al., 2008). As the gametophytes of kelp species are microscopic and thus cannot be studied in the field due to technical constraints, our focus has been made on the sporophyte stage of *L. digitata*. Sporophytes of *L. digitata* were sampled in four locations spanning from the center to the southern margin of its north-eastern Atlantic distribution. The sampling sites included Clachan, Scotland (CLA); Helgoland, Germany (HLG); Roscoff, France (ROS) and Quiberon, France (QUI) (Table 1, Figure 1). Tissue samples of approximately 3 cm were collected from 30-40 sporophytes in the low intertidal zone at each site and were promptly stored in silica gel to preserve their genetic material. For three out of the four populations (HLG, ROS, QUI), samples were collected at two time points from the same site to capture changes in allele frequencies over time and assess the relative influence of local selection and genetic drift. Given the lifespan of sporophytes and the interval between samplings, temporal samples are separated by at least two generations. The first set of samples was collected between 2005 and 2011, while the second set was collected in 2018.

Regarding the CLA population, samples were collected in 2008 and 2018 from two separate sites, CLA_1 and CLA_2, which were approximately 20 km apart (Figure 1). Given the frequent observation of genetic structure in this species at a scale of 1-10 km (e.g., Billot et al., 2003; Robuchon et al., 2014), samples from CLA were treated as separate populations rather than two time points of the same population.

## 2.2 Temperature data

Given the increasing evidence that temperature has the potential to trigger adaptive differentiation among kelp populations (e.g., King et al., 2019; Liesner et al., 2020; Schimpf et al., 2022) sea surface temperature (SST) appeared as the most relevant selective agent for assessing adaptive differentiation in *L. digitata*. Daily mean SST were obtained from satellite observations spanning 37 years (1981-2018), with a spatial resolution of 0.05° × 0.05° for QUI and 0.02° × 0.02° for the other sites (sourced from E.U. Copernicus Marine Service, 2022). This long-term period was chosen to evaluate the trend and intensity of climate warming in the studied populations and relate it to the contemporary adaptive response. Warming was evaluated by performing a linear regression between time (measured in years) and average SST of the warmest months (June, July and August) for each site. The PELT (Pruned Exact Linear Time) algorithm, implemented in the changepoint v2.2.4 R package, was subsequently employed to identify significant shifts in SST for each population and delimit two thermal periods: one before and one after the significant shift. For each period, we subsequently defined the monthly average SST of the warmest months (referred to as 'highest mean SST' hereafter), of the coolest months (i.e., from January to March, and referred to as 'lowest mean SST' hereafter) and across the seasons ('mean SST').

## 2.3 Library preparation, genotype and SNP calling

DNA extraction was performed on dried tissue preserved in silica gel from three to 16 years using the Nucleospin R 96 plant kit II (Macherey-Nagel, Düren, Germany), following the manufacturer's protocol. Two double-digest RAD-sequencing libraries (ddRAD-seq, Peterson et al., 2012) were prepared, with 117 individuals for the first time point and 104 individuals for the second time point, including eight replicates (i.e., eight samples with the same DNA extract but independent library preparation, sequencing, read mapping, and SNP calling) in each population and time point. Individuals were randomly distributed across libraries, along with 355 samples of distinct projects to prevent batch effects and ensure library diversity. Library preparation was conducted according to Reynes et al., (2021), using 100 ng of DNA and the *PstI* and *HhaI* enzymes (NEB). Paired-end 150 reads were obtained by sequencing the libraries on an Illumina Hiseq 4000 platform (Génome Québec Innovation Centre, McGill Univ., Montreal, Canada). Quality control of raw reads and adaptor removal was performed using FASTQC v.0.11.7 (Andrews, 2010) and Trimmomatic (Bolger et al., 2014), respectively.

Demultiplexing was carried out using *process_radtags* with default parameters, implemented in the Stacks v.2.52 pipeline. Individual paired-end reads were trimmed to 137 bp using Trimmomatic (Bolger et al., 2014) and mapped to the draft genome of *L. digitata* (Dryad Digital Repository, see Data Availability Statement) using BWA-mem with default parameters in BWA v.0.7.17 (Li & Durbin, 2009). The N50 of the genome is 9.9 Mb with a genome assembly size of ~470 Mb among 671 555 scaffolds. Uniquely mapped reads were retained with SAMtools v.1.13 (Danecek et al., 2021). Generated BAM files were processed for SNP calling using the reference mode of the Stacks v.2.52 pipeline (Catchen et al., 2011, 2013; Rochette et al., 2019). Genotyping and SNP calling were carried out separately for each time point, and the shared SNPs between temporal datasets were selected before merging individuals across these SNP positions. The *bcftools isec* and *bcftools merge* functions of BCFtools v.1.9 (Danecek et al., 2021) were used for these steps. Post-call filtering of SNPs was performed on the merged dataset by keeping SNPs with a call rate >90% per population in one or more populations using *pop_missing_filter.sh* of the dDocent pipeline (Puritz et al., 2014). Problematic individuals (n = 23) having a call rate below 80% were discarded. Filtering based on mean read depth (DP) and minor allele frequency was executed using vcftools v.0.1.16 (Danecek et al., 2011), with specific parameters indicated in Table 2. To address the effects of excessive linkage disequilibrium (LD) between loci, SNPs with a square correlation ($r^2$) > 0.2 were removed using PLINK v.1.9 (Chang et al., 2015). The calculation of $r^2$ values was performed separately for each population to separate the effects of physical proximity among SNPs from the effects of population structure in LD patterns. SNPs with $r^2$ values exceeding the threshold in all populations were discarded (Table 2). After all post-filtering steps, a total of 2 854 SNPs remained among 190 individuals (excluding the eight replicates). Genotyping concordance was assessed in this dataset by calculating the SNP error rate between replicate pairs following Mastretta-Yanes et al., (2015).

## 2.4    Genetic differentiation

An analysis of molecular variance (AMOVA) implemented in Arlequin v.3.11 (Excoffier et al., 2005), with 10 000 permutations was performed to test for temporal changes. We tested for changes in genetic variation between two time periods for each of the three populations that were resampled over time (HLG, ROS, QUI). Spatial and temporal genetic differentiation was quantified by calculating $F_{ST}$ (Weir & Cockerham, 1984) using the R package diveRsity v.1.9.90 (Keenan et al., 2013). Genic differentiation was tested with an exact test in GENEPOP v.4.7.5 R package (Raymond & Rousset, 1995) with the Markov chain method and default parameters. A combination of all tests across loci was performed using Fisher's method. Genetic structure was investigated using the sNMF (sparse Non-Negative Matrix Factorization) algorithm implemented in the R package LEA v.2.8 (Frichot et al., 2014; Frichot & François, 2015). The analyses involved 10 000 iterations and 20 repetitions, with K ranging from 1 to 16. The best K value was determined based on the cross-entropy criterion. A complementary analysis of genetic structure was performed using a principal component analysis (PCA) on SNP variation, with the R packages Bigsnpr v.1.3.0 and Bigstatsr v.1.2.3 (Privé et al., 2018). Missing genotypes were imputed by replacing them with the average allele frequencies before performing the PCA.

## 2.5    Genetic diversity and effective population size

Genetic diversity within populations was estimated using the complete set of 2 854 SNPs. Expected heterozygosity (He), and observed heterozygosity (Ho) were estimated using the R package diveRsity. Differences in He between populations and time points were tested using pairwise Wilcoxon tests, with adjustment for multiple comparisons using the Bonferroni method. The proportion of polymorphic loci (P%) within each population and time point, has been computed with a custom bash script. A confidence interval on P% was calculated by randomly sampling 20 individuals with replacement within populations, and iterating this procedure 100 times. The extent of local genetic drift in temporal differentiation was additionally assessed by investigating whether the minor allele was retained over time given its frequency at the initial time point.

The contemporary effective population size (Ne) was estimated using temporal variance in allele frequencies with three methods: two methods based on F-statistics, $F_c$ (Nei & Tajima, 1981) and $F_s$ (Jorde & Ryman, 2007), and a likelihood estimator (Wang & Whitlock, 2003), known for its improved precision and accuracy in the presence of rare alleles (Wang, 2001). The software Neestimator v.2.1 (Do et al., 2014) was used for methods based on F-statistics and MLNe v.2.0 (Wang & Whitlock, 2003) for the likelihood approach. The 95% confidence intervals of the likelihood estimate and moment of the F-statistics were also calculated. Ne estimations were performed considering the plan II sampling procedure, which assumes that individuals are sampled before reproduction and are not returning to the population (Waples, 1989). Finally, given the *L. digitata* lifespan and the interval between samplings, we assumed that two generations had passed between the time points.  Ne was estimated twice: first, including both neutral and outlier SNPs, and second, by excluding spatial outliers detected by at least two genome scan methods. These spatial outliers are more likely to be influenced by strong directional selection.

## 2.6    Outlier detection across space

Outlier tests based on spatial differentiation were conducted to identify loci exhibiting high spatial differentiation, potentially deviating from neutral expectations, and indicating the possible influence of local selection. These outlier tests were performed separately for each time point and included individuals from CLA_1 (2008) and CLA_2 (2018). Three different methods were employed for outlier detection across space. We first used the Bayesian approach of Beaumont & Balding (2004) implemented in BayeScan 2.1 (Foll & Gaggiotti, 2008). The program was run with different prior odds (3, 5, 10 and 100) with 20 pilot runs of 5 000 iterations each, followed by a burn-in of 50 000 iterations and 5 000 samplings. We then ran a PCA using the R package pcadapt (Luu et al., 2017) to assess the contribution of each SNP to the K principal components (PCs). Lastly, we used the OutFLANK v.0.2 software (Whitlock & Lotterhos, 2015) to identify outliers by comparing differentiation at each SNP against a trimmed null distribution of $F_{ST}$ values. OutFLANK was run with LeftTrimFraction = 0.55 and RightTrimFraction = 0.10. To account for multiple testing, the p-values obtained from each method were corrected using the R package qvalue v.2.18, with a threshold set at 0.10. Overlapping outlier SNPs across methods and datasets were analyzed using jvenn (Bardou et al., 2014).

## 2.7    Outlier detection across time

Outlier tests were conducted between time points to distinguish the effects of selection from genetic drift in allele frequency changes. First, we simulated SNP frequencies over two generations using SLiM 3 (Haller & Messer, 2019) to assess whether the observed patterns of temporal differentiation are consistent with those expected by genetic drift. For each simulated SNP, the frequency estimated in the population at the first time point was used as the SNP frequency at the beginning of the simulation. For each population, the simulations were iterated 5 000 times. We assumed panmictic reproduction and neglected mutation and migration. The simulations lasted two generations, according to the aforementioned time interval between sampling points. For each of the three populations (HLG, ROS, QUI), simulations were conducted using the lowest and highest estimates of Ne based on temporal methods and by assuming constant population size over time.  At the end of simulations, N individuals were sampled, with N corresponding to the sampling size reported (see Table 1). This framework served as a baseline for detecting candidate loci undergoing directional selection, as they would exhibit higher temporal differentiation than expected under neutral evolution. The vcf output files including empirical and simulated SNPs (5 000 files for each population and Ne scenario) were processed using vcftools to calculate pairwise $F_{ST}$. Finally, the p-value of the outlier test at a focal SNP was computed as the proportion of simulated $F_{ST}$ that was equal to or larger than the observed $F_{ST}$. This was performed using a custom R script, which is accessible at https://lauricreynes.github.io/Temporal-genomics/.  In the main manuscript we will only show the results obtained from the outlier test ran on the high Ne scenario which is expected to underestimate the level of drift. However, the results were similar when using the low Ne scenario (Supplementary S1). We additionally compared our simulation to the method implemented in TempoDiff, which also aims to distinguish neutral from selected polymorphisms using temporal differentiation (Frachon et al., 2017). We also ran BayeScan, OutFLANK and pcadapt using the same parameters as reported in the section 'Outlier detection across space'.

The p-values of the tests were corrected for multiple testing before conducting an overlap analysis among methods, as described in the previous section.

## 2.8    SNPs-temperature associations

The association between SNPs and temperature was investigated across space using a logistic regression framework, focusing specifically on SNPs identified as outliers by at least two tests across different geographical locations (see previous section 'Outlier detection across space'). In the context of strong directional selection, logistic regression best meets the assumptions underlying a sigmoidal pattern compared to linear regression (Rellstab et al., 2015). Logistic regression was applied to the presence or absence of the alternative allele, respectively coded as 0 or 1 for individual genotypes. As we aimed to detect signals of local selection across populations, the analysis was conducted using the complete set of individual genotypes, associating them with SST parameters corresponding to those estimated for each period (T1 and T2; see Table 3). SST predictors included the 'highest mean SST', 'lowest mean SST' and 'mean SST'. To account for potential false positives stemming from shared ancestry, the first five PCs of SNP variation were incorporated as covariates in the logistic regression model. The logistic regression was carried out using the glm R function, specifying binomial variance and a logit link function. To refine the model and select the most relevant predictors, a stepwise variable selection procedure based on the Akaike Information Criterion (AIC) was performed in both forward and reverse directions using the stepAIC function from the MASS R package. The fitting of the model was evaluated using McFadden's pseudo R-squared, a commonly used metric for assessing the goodness of fit. Given that logistic regression was performed 'n' times, corresponding to the number of outlier SNPs, a strict Bonferroni correction was applied to adjust for multiple testing ($P < 0.05/n$). Only models for which at least one SST predictor showed a significant p-value after correction were retained. From the coefficients (slopes) of the fitted models, odds ratios and confidence intervals were estimated for each predictor. The fitted model's predictions were then visualized and compared to the observed data.

## 2.9    Gene ontology analysis

Candidate loci identified through outlier tests across space and time by at least two methods were subjected to functional annotation using Omics Box v.1.3.11 (Götz et al., 2008). Initially, candidate loci were annotated using the NCBI Basic Local Alignment Search Tool (BLAST, Johnson et al., 2008) with the non-redundant protein sequences database. The BLASTx approach was employed, with a specific focus on Phaeophyceae sequences. The resulting BLAST hits were further mapped using InterProScan (Zdobnov & Apweiler, 2001) and Gene Ontology (Ashburner et al., 2000). Finally, both analyses were merged to obtain comprehensive functional annotation for candidate loci.

## 3. RESULTS

### 3.1    Warming trends

Sea Surface Temperature (SST) recorded between 1981 and 2018 displayed noticeable variations among the investigated populations. The 'highest mean SST' varied from 13°C in CLA to 17.7°C in HLG and QUI (Table 3), and the 'lowest mean SST' was the lowest in HLG (around 4°C), indicating that the latter population experienced the greatest variation over a season. Populations in Brittany experienced the most temperate winter period, with SST never dropping below 9.3°C (Table 3). In each studied population, a substantial rise in SST was observed over 37 years, illustrating the impact of global warming in that area. Rise in SST was particularly notable in HLG, in which the 'highest mean SST' increased by 0.048°C / year according to the linear regression (estimate = 0.048, sd = 0.009, p-value < 0.0001, $R^2$adj = 0.18, Figure 2). This trend contrasted with those reported in Brittany, for which the model estimated an average increase of 0.013°C / year in ROS (estimate = 0.013, sd = 0.004, p-value = 0.004, $R^2$adj = 0.07) and 0.009°C / year in QUI (estimate = 0.009, sd = 0.004, p-value = 0.035, $R^2$adj = 0.03), respectively. Given that SST increase was variable among the studied populations, the Pruned Exact Linear Time (PELT) method identified distinct changepoints when splitting the 37 years into two. These changepoints varied across populations between the years 1993 (QUI) and 2004 (HLG) (Table 3, see period).

### 3.2    Sequencing, SNP filtering and data quality

A total of 98.7 million reads were obtained for the first time point and 327 million reads for the second time point. The individual proportion of mapped reads was on average 86.6% for the first time point and 95.5% for the second time point. Both the count of high-quality reads and the mapping rate were significantly lower in the first time point compared to the second time-point (Wilcoxon test, p-value < 0.001). As a result, SNPs were approximately 8.3 times higher for the second time point (644 794 SNPs) compared to the first time point (77 193 SNPs). After filtering for shared loci between time points, with a call rate >90% per population in one or more populations, a total of 4 151 SNPs was retained (Table 2). Quality filtering, including individual missingness, read depth, minor allele frequency (MAF), and linkage disequilibrium further refined the dataset to a final set of 2 854 SNPs across 190 individuals, excluding eight technical replicates (Table 2). The analysis of the eight replicates indicated a high level of genotype concordance for the set of 2 854 SNPs, which was consistent across both time points. The SNP error rate ranged from 0.007 to 0.036 for the first time point and from 0 to 0.02 for the second time point (Table S1). The mean read depth was 16.7 X ± 17.9 SE for the first time point and 22.5 X ± 14.2 SE for the second. Despite slight differences in sequencing depth between the two time points, we found no significant difference in individual heterozygosity (Wilcoxon test, p-value = 0.23).

### 3.3 Genome-wide genetic diversity

The expected heterozygosity (He) was significantly different between sites (Wilcoxon test, p-value < 0.001). The northern populations (CLA_1 and CLA_2) showed the highest heterozygosity (He = 0.103 and He = 0.109 for CLA_1 and CLA_2, respectively), which was seven times higher than for HLG (He = 0.015 and He = 0.013 for T1 and T2, respectively). Furthermore, He values in CLA_1 and CLA_2 were 1.5 to 2.7 times higher compared to those observed in Brittany (ROS and QUI, respectively, Table 1). Regarding genetic diversity over time, He slightly decreased (~2%) in HLG (Wilcoxon test over SNPs, p-value < 0.001), while its slight increase in Brittany was statistically significant for QUI (p-value = 0.047) but not for ROS (p-value = 0.13). The decline (~3%) in the proportion of polymorphic loci was only reported in HLG (Table 1), further supporting its overall decrease in genetic diversity. Regarding, allelic variation over time, the proportion of minor alleles (MAF < 0.05) not detected at the second time point differed across populations: HLG (0.69), ROS (0.27) and QUI (0.47) (Figure 3). In addition, the proportion of alleles occurring at higher frequencies (MAF $\geqslant$ 0.05) at the first time point and not detected at the second was more pronounced for HLG (0.27) than for ROS (0.03) and QUI (0.10).

### 3.4 Genetic structure

The AMOVA analysis conducted using the complete SNPs dataset revealed that variation was primarily attributable to variance within individuals (81%, p-value < 0.001) and secondly between populations (21%, p-value=0.002), while the effect of time was not significant (-0.61%, p-value = 0.708, Table 4). The average $F_{ST}$ values among populations ranged from 0.196 to 0.221 for the first and second time point, respectively. When excluding CLA_1 and CLA_2, the average $F_{ST}$ values were 0.100 and 0.129 for the first and second time points, respectively. All pairwise $F_{ST}$ values in space were greater than or equal to 0.078 with a maximum of 0.297 observed between CLA_2 and the second time point of HLG (Table S2). All pairwise tests of genic differentiation for spatial comparisons were significant (p-values < 0.001), while none of the tests were significant for temporal comparisons (p-values = 1). This latter result is consistent with the fact that pairwise temporal $F_{ST}$ values were close to zero (Table S2). The lack of significant differentiation over time was further confirmed by the Sparse Non-negative Matrix Factorization (sNMF), as there was no grouping by date and no important change in genetic structure between time points (Figure S1). When considering the most informative number of clusters according to the minimum cross-entropy criterion (K = 5), individuals were grouped by sites, with an additional distinction between CLA_1 and CLA_2.

### 3.5 Candidate SNPs for local selection in space

Among the 261 outliers (9.1 % of the total) identified by at least one method, 97 (3.4% of the total) were shared between time points, illustrating that genetic differentiation at these SNPs was conserved over time. On the other hand, 47 SNPs were exclusively detected in the first time point, and 88 SNPs in the second. Interestingly, 36 of the SNPs detected in space were also identified as outliers by the temporal data-based tests. To reduce the number of false positives, we further considered only those SNPs that were detected by at least two out of the three detection methods in space. By doing so, none of the 47 outlier SNP remained for the first time point, while 16 out of the 88 outlier SNPs remained for the second. Among these 16 SNPs, 15 were detected by both OutFLANK and pcadapt, and only one SNP was detected by all three methods (OutFLANK, pcadapt and BayeScan).

### 3.6    Ne is closely related to temporal outliers

When searching for temporal outliers, 149 SNPs (5,2% of the total) were identified by at least one method: TempoDiff, OutFLANK, or the simulation framework. BayeScan and pcadapt did not detect any outliers. The outlier tests based on our simulation framework reported the occurrence of 131 SNPs with $F_{ST}$ values higher than the neutral expectations (Figure 4). Among the 53 SNPs detected by TempoDiff, 35 were identified by the simulation framework, including 13 SNPs that were detected by all three methods. When solely considering the SNPs detected by at least two methods, the number of outlier SNPs decreased to 38, among which 23 were detected in ROS, 15 in QUI and none was detected in HLG (Figure 5). Interestingly, the absence of temporal outliers was observed in the population with the lowest effective size. Temporal estimates of Ne indicated that the HLG population had the smallest Ne values compared to those reported at ROS and QUI. Although Ne estimates varied depending on the estimator used, and whether spatial outliers were considered in the estimation of Ne or not (see Table 5), all estimators followed this trend and underscored the small effective size of the HLG population. Considering observed heterozygosity (Ho) for outlier SNPs reported in ROS and QUI, Ho increased over time from 0.125 to 0.192 for ROS and from 0.087 to 0.184 for QUI. Further analyses of allele frequencies indicated that the increase in Ho was primarily attributed to an increase in low-frequency variants between time points. Specifically, the Minor Allele Frequency (MAF) increased on average from 0.07 to 0.23 in QUI and from 0.07 to 0.14 in ROS (Figure S3). This pattern starkly contrasted with the low Ho values observed in HLG, which decreased from 0.033 to 0.021 over time.

### 3.7    SNPs-temperature associations

We conducted a logistic regression to analyze the occurrence of the alternative allele as a function of sea surface temperature (SST) predictors, for the 16 SNPs identified as outliers in the spatial analysis. This analysis indicated that three of these 16 SNPs were correlated with SST predictors. However, after applying correction for multiple testing, two of these remained statistically significant (P<0.003) and correlated with the 'highest mean SST'. Among the principal components (PCs) of genetic variation, only PC1 remained statistically significant after applying correction for multiple testing. PC1 was associated with three SNPs out of the 16 that were different from the ones correlated with SST predictors. The 10 remaining SNPs did not show any significant associations with either the PCs of genetic variation or the SST predictors. A summary of the model with predictor significance after conducting a stepwise variable selection procedure is presented in Table S3. The GLM for the two SNPs correlated to the 'highest mean SST', predicted an increase in the occurrence of the alternative alleles as the 'highest mean SST' increased (Figure 6). This aligns with the expectation in cases where directional selection favors different alleles in different populations. However, the response curves of the GLMs are relatively contrasted between these SNPs, suggesting that SST may not have the same effect on their alternative alleles. For instance, the response curve of the SNP ID '144563:157' indicated that an increase of one unit in the 'highest mean SST' raised the probability of the alternative allele's presence by 8.03 (OR = 8.03; 95% CI: 4.42-21.7; P < 0.003). This is supported by empirical observations, which highlight that the alternative allele at this SNP is nearly absent when the 'highest mean SST' is lower than 15°C, as observed in CLA (Figure 6), and increases in frequency until reaching fixation at 17.5°C, as observed in QUI. In contrast, the response curve of the second SNP (ID '71212:99'), indicated that both reference and alternative alleles are predicted to co-occur in equal frequency at intermediate temperatures, gradually increasing or decreasing in frequency with rising or falling temperatures (Figure 6).

### 3.8    Gene ontology

Functional analyses were conducted using the OmicsBox 2.2.4/Blast2GO pipeline, focusing on 53 loci with SNPs detected as outliers in space (16 SNPs) and time (38 SNPs), identified by at least two methods. Out of these 53 loci, eight showed significant hits in the BLASTX search against the UniProt database. The top hits were associated with brown algae of the *Ectocarpus* genus, featuring e-values ranging from 2.8E-4 to 6.4E-19. Six of the targeted loci were annotated to diverse Gene Ontology terms (see Table S4). In the context of spatial outlier SNPs, particularly the locus with ID '71212', which demonstrated correlation with the 'highest mean SST' in SNPs-temperature associations, functional annotation revealed associations with domains involved in lipid transport and cellular anatomical entity. The remaining three annotated loci associated with spatial outlier SNPs were linked to functions involving membrane structures, oxidoreductase activity, and methyltransferase activity. Regarding the three annotated loci associated with temporal outlier SNPs, the analysis indicated specific functions ranging from phosphorylation and kinase activity to mismatch repair and RNA polyadenylation (Table S4).

# 4. DISCUSSION

We studied the contemporary evolution of populations of the kelp *Laminaria digitata* by examining spatial and temporal genetic variation. While the detection of outlier SNPs through space, partly linked to Sea Surface Temperatures (SST), pointed out that temperature may drive adaptive response, temporal genomics offered further insights into the efficiency of selection relative to genetic drift. With the support of these spatial and temporal analyses, we delve into a detailed discussion regarding how the studied populations are currently responding to the challenges posed by climate change.

## 4.1 Adaptive potential at the southern margin

Analyses of changes in SST indicate that the studied populations experienced distinct thermal regimes that gradually became warmer over the study period. Nevertheless, our findings revealed variations in warming trends across populations, suggesting that vulnerability to climate change and selection for climate-related traits may not occur uniformly throughout the species' geographical range. While our results emphasized the moderate effective size (Ne) of the southernmost population (Quiberon) compared to the core one (Roscoff), we expected that selection efficiency would be reduced at the southern margin. This hypothesis aligns with theoretical expectations and is corroborated by previous studies on *L. digitata*, which reported low within-population variation for Quiberon using microsatellites (Oppliger et al., 2014; Robuchon et al., 2014; Liesner et al., 2020).

Even though Ne was smaller for Quiberon than Roscoff, we did not detect any evidence of reduced adaptive potential in Quiberon. This is supported by the detection of two types of candidate loci: i) loci exhibiting high spatial differentiation and potentially associated with SST, and ii) loci showing greater temporal differentiation than expected under genetic drift alone. Additionally, some of these candidate loci were shown to be linked to metabolic and cellular functions that might be involved in adaptive responses, such as variations in thermal regimes. Our findings, therefore, support that SST may drive the adaptive response of *L. digitata*, in line with previous heat stress experiments conducted on the same populations (Liesner et al., 2020; Schimpf et al., 2022) and genome scans from other kelp species (Guzinski et al., 2020; Mao et al., 2020; Vranken et al., 2021). Nonetheless, confounding factors may also be at play, such as the phylogeographical history of the species along the north-eastern coast of the Atlantic which would split the studied populations into two different clusters (Neiva et al., 2020). In this context where genetic structure complicates the inference of selective loci (see Bierne et al., 2013), our temporal genomics framework, applied over a short time interval, offered further insights on the detection of non-neutral loci.

The identification of temporal outliers over a short period, particularly during a period marked by climate warming, raises questions about the role of SST in driving temporal differentiation. Nonetheless, despite a moderate increase in SST over the study period at Quiberon, temporal outliers were mainly observed in this population. Several hypotheses can be advanced to explain this observation: (i) Other selective factors besides SST may be influencing the observed temporal differentiation. (ii) Fine-scale variations in temperature, especially in coastal areas, may not be adequately captured by the available SST data. (iii) Genetic differentiation could potentially be a reflection of significant changes in SST that occurred before the study period. Regarding the latter assumption, genetic differentiation at Quiberon may indeed reflect a gradual response to selective pressures imposed by climate change over successive generations. This aligns with the theory that populations' adaptation lags (see Martin et al., 2013 for a review) increase with marginality while stronger intensity is predicted at the species' warm margin (Hampe & Petit, 2005; Fréjaville et al., 2020).

## 4.2    The role of genetic drift in shaping genetic variation

The study of the population from the island of Helgoland gave contrasting results compared to Quiberon. Similar climate-related selective pressures were expected in Quiberon and Helgoland as both populations experienced warmer SST (Derrien-Courtel et al., 2013) and were shown to exhibit slightly higher thermal tolerance than populations from the cool margins (Liesner et al., 2020). However, Helgoland exhibited the lowest level of genome-wide genetic diversity among the populations included in this study and its diversity tended to decrease over time. In addition, candidate loci with elevated temporal differentiation were rare in Helgoland, even absent when considering those detected by at least two genome scan methods. The high environmental variability observed in this population may have led to increased plasticity for a wide range of temperatures, thereby reducing the selection pressure and explaining the rarity of temporal outliers. Another explanation is the detrimental effect of accelerated warming, surpassing that of the other sites, and leading to major demographic reduction. The high incidence of genetic drift in this population was evidenced by a reduction in the number of polymorphic loci over time, mainly marked by the loss of low-frequency alleles. This is in line with earlier studies which have shown low levels of genetic diversity at microsatellites loci (Liesner et al., 2020; Fouqueau, 2021) and patterns of local adaptation to temperature that became less clear when including gametophytes from Helgoland rather than Quiberon (Schimpf et al., 2022). This genetic impoverishment likely results from a smaller effective population size and/or geographical isolation (Blows & Hoffmann, 1993; Johannesson & André, 2006; Eckert et al., 2008; Pilczynska et al., 2019), potentially exacerbated by climate change. For instance, Bartsch et al., (2013) noted that the prevailing SST of approximately 18°C in August inhibited the summer reproduction of the species in Helgoland. Furthermore, climate change appears to have played a major role in the substantial biomass decline of *L. digitata* (in terms of sporophytes) observed between 1970 and 2005 in Helgoland, as reported by Pehlke and Bartsch (2008). The demographic and genetic erosion observed here does not appear to be limited to *L. digitata* alone. The kelp *Saccharina latissima* also experienced low levels of genetic diversity, as evidenced by both microsatellites and

SNPs (Guzinski et al., 2020). The contemporary response of kelp populations in Helgoland could serve as a valuable model for gaining insight into how kelp populations might respond to climate change, particularly for small and fragmented populations already damaged by high rates of genetic drift (e.g., Arizmendi-Mejía et al., 2015; Crisci et al., 2017).

## 4.3 Technical considerations

DNA damage can be a concern in temporal genomics studies, particularly when they have been carried out on historical samples (Dehasque et al., 2020; Therkildsen et al., 2013a, b). In our study, the decrease in the number of reads and SNPs in the oldest samples compared with the most recent ones may be the result of lower DNA quality for the oldest time points. As shown by our results, this implies that only a small proportion of the loci sequenced at the second time point was recovered at the first time point. Although this reduces the amount of data to study temporal genetic changes at putatively selected loci, our filtering strategy, by retaining SNPs shared between time points with enough allele depths, together with an analysis of genotype concordance at each time point, allows us to conclude that DNA damage is not relevant here to explain temporal genetic changes. An important aspect of this study is that the earliest sampling point did not occur in the same year for all populations due to practical constraints. We expect to enhance our ability to assess the effects of climate change on genetic variation by increasing the interval between sampling periods. In contrast, reducing the interval between samplings or delaying the initial sampling until after significant climatic changes have occurred may limit such investigations, as the genetic makeup of the sampled generations could already be influenced by climate change. However, our results highlighted that temporal outlier SNPs were detected in Roscoff and Quiberon, even though both populations had the shortest interval between samplings. This underlines that temporal genomics spanning two or three generations can be highly valuable in detecting ongoing selection, especially in species with short generation times.

## 4.4 Beyond selection: factors influencing temporal outliers

By investigating genetic variation among the sporophyte stage, our study only captures half of the species' biphasic life cycle. Gametophytes have the potential to persist as a genetic bank of microscopic forms (e.g., Edwards, 2000; Bartsch et al., 2013) and are generally associated with higher thermal resistance than sporophytes (Bolton & Lüning, 1982). Both of these characteristics suggest that gametophytes may serve as a buffer, retaining genetic variation when episodes of heatwave cause a severe bottleneck in the sporophytic population. Although the importance of gametophytes in preserving the evolutionary potential is still poorly understood (see Veenhof et al., 2022 for a review), our results rather question the idea that gametophyte stages represent important hidden genetic variance. Indeed, if the population of Helgoland has been replenished from its microscopic forms as suggested by Bartsch et al., (2013), the observed low genomic variation suggests that a restricted number of gametophytes might have contributed.

Finally, it is important to discuss the assumptions that were made for estimating contemporary Ne and detecting temporal outliers. Firstly, we assumed that temporal samples were separated by two sporophytic generations, whichever the populations, although life spans of *L. hyperborea* sporophytes are shortened at warmer temperatures (Bartsch et al., 2008). Yet when considering three generations rather than two, the effect of genetic drift might have been underestimated. This is especially of concern for the southernmost population which is confronted with the warmest climate. However, given that loci detected by at least two methods of temporal genome scans exhibited $F_{ST}$ values exceeding 2.4 to 3 times the threshold of neutral expectation, this effect should be minimized here. Secondly, we made the assumption of isolated populations for estimating Ne. However, signatures of elevated temporal differentiation may be expected in the presence of substantial levels of gene flow (e.g., Therkildsen et al., 2013a, b). While the effect of gene flow from unsampled populations in the detection of temporal outliers cannot be ruled out, gene flow is still expected to be extremely limited among local populations (Fouqueau et al., 2023). Gene flow is anticipated to exert substantial influences on allele frequencies within species characterized by genetic structure operating at the scale of a few kilometers. While our results indicate that temporal differentiation was not significant when considering the full set of SNPs, this does not rule out the possibility that there have been moderate changes in allele frequencies due to local gene flow. Further analysis using chromosome-wide assembly will be necessary to determine whether temporal outlier SNPs are clustered in the same genomic regions or distributed randomly across distinct regions. While careful interpretation is essential regarding temporal genetic changes, selection remains the most pertinent factor for explaining the detection of temporal outliers in *L. digitata* populations."

## 4.5    Conclusion

Our spatio-temporal study on *Laminaria digitata* populations has provided valuable insights into the species' contemporary evolution. We have documented contrasting evolutionary responses, highlighting the variability in how populations react to the challenges imposed by climate change. This understanding has been significantly bolstered by temporal genomics, which provided a more detailed view of the interplay between genetic drift and selection in contemporary times. Given the high threat that climate warming poses to kelps and other marine foundation species, we anticipate that spatio-temporal genomic frameworks will become more common, encouraging monitoring and guiding conservation efforts.

**Acknowledgements**

This work is especially dedicated to the memory of Gernot Glöckner who contributed to the sequencing of *Laminaria digitata* genome and passed away in very recent time. This work was funded by the EU project MARFOR Biodiversa/004/2015. LF was additionally funded by the Region Bretagne (ARED 2017 REEALG) and the NOMIS foundation. The authors thank the ABiMS platform of the Roscoff biological station (http://abims.sb-roscoff.fr) for providing the HPC resources that contributed to the search results reported in this document. We also acknowledge the staff of the "Cluster de calcul intensif HPC" Platform of the OSU Institut Pythéas (Aix-Marseille Université, INSU-CNRS) for providing the computing facilities. The project leading to this publication has received funding from EC2CO (CNRS) fund and from European FEDER Fund under project 1166-39417.

**Competing interests**

The authors declare no competing interests.

**Data Availability Statement**

Individual high-quality reads (paired-end reads) generated in the present study (fq.gz files) and the filtered VCF file can be accessed via the Dryad repository at https://doi.org/10.5061/dryad.tb2rbp064. Additionally, the scaffolds assembly of the *Laminaria digitata* genome, utilised for read mapping and produced by Gernot Glöckner, is also available in the same repository. Furthermore, the custom R script developed for identifying temporal outlier SNPs, along with the $F_{ST}$ datasets comparing observed and simulated $F_{ST}$ between time points, can be found on GitHub at https://doi.org/10.5281/zenodo.10946282.

**Figure 1.** Sampling sites of the *Laminaria digitata* populations used in this study, with populations from Clachan highlighted in a zoomed-in area at the bottom right corner.

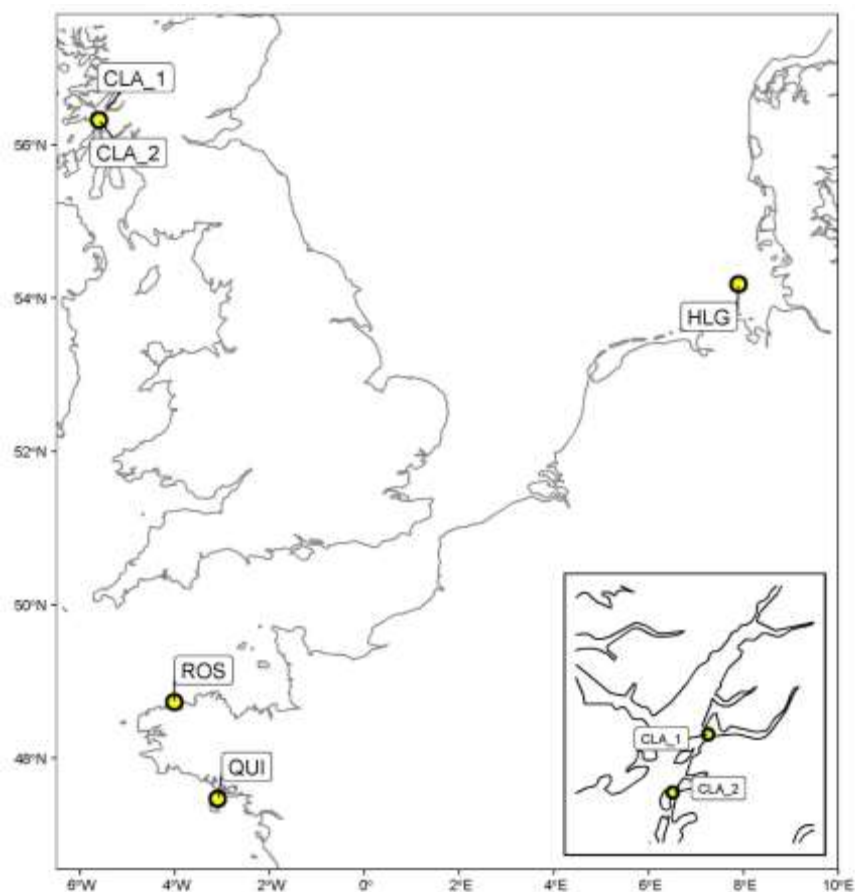

**Figure 2.** Linear regression analysis between the average Sea Surface Temperature (SST) of the warmest months ('highest mean SST') and the Years. Each population was subject to separate linear regression analysis with the corresponding regression equation displayed in the bottom right corner of each subplot. Note that the Y-axis scale is not uniform throughout the entire graph.

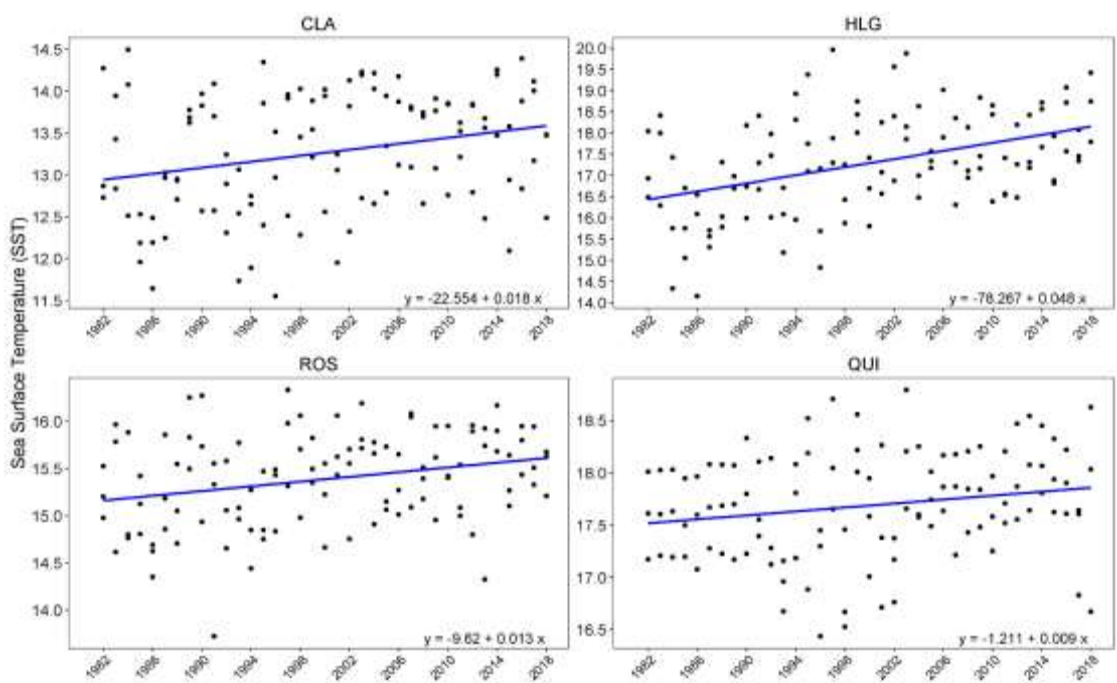

**Figure 3.** Minor allele frequencies (MAF) change between time points. Alleles not reported at the second time point are depicted in red, while retained polymorphic shown in blue.

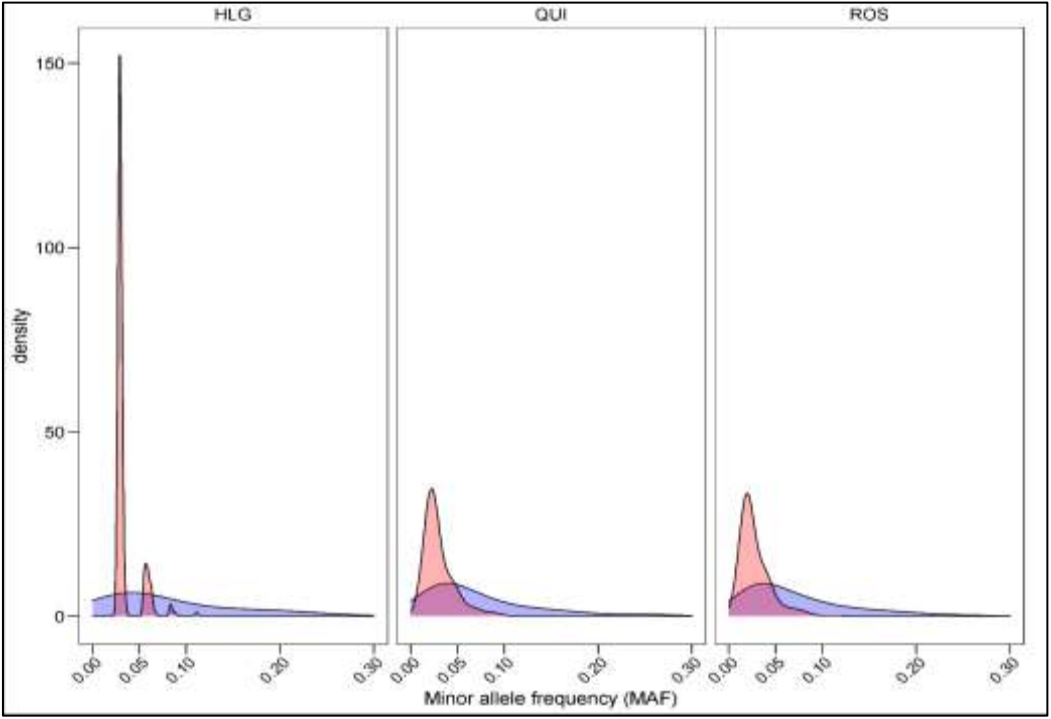

**Figure 4.** Temporal $F_{ST}$ per SNP is compared to the upper limit of the neutral expectation. The green line represents the 99% quantile of $F_{ST}$ values expected under genetic drift alone, determined from 5 000 simulations for (A) HLG, (B) ROS, and (C) QUI. Outlier SNPs detected only by the simulation framework are represented by gray triangles. Those detected by two or three methods are represented by purple and red triangles, respectively. Gray crosses indicate putatively neutral SNPs that were not detected by outlier tests. Note that the order of SNPs is based on contigs and does not reflect their position along the genome.

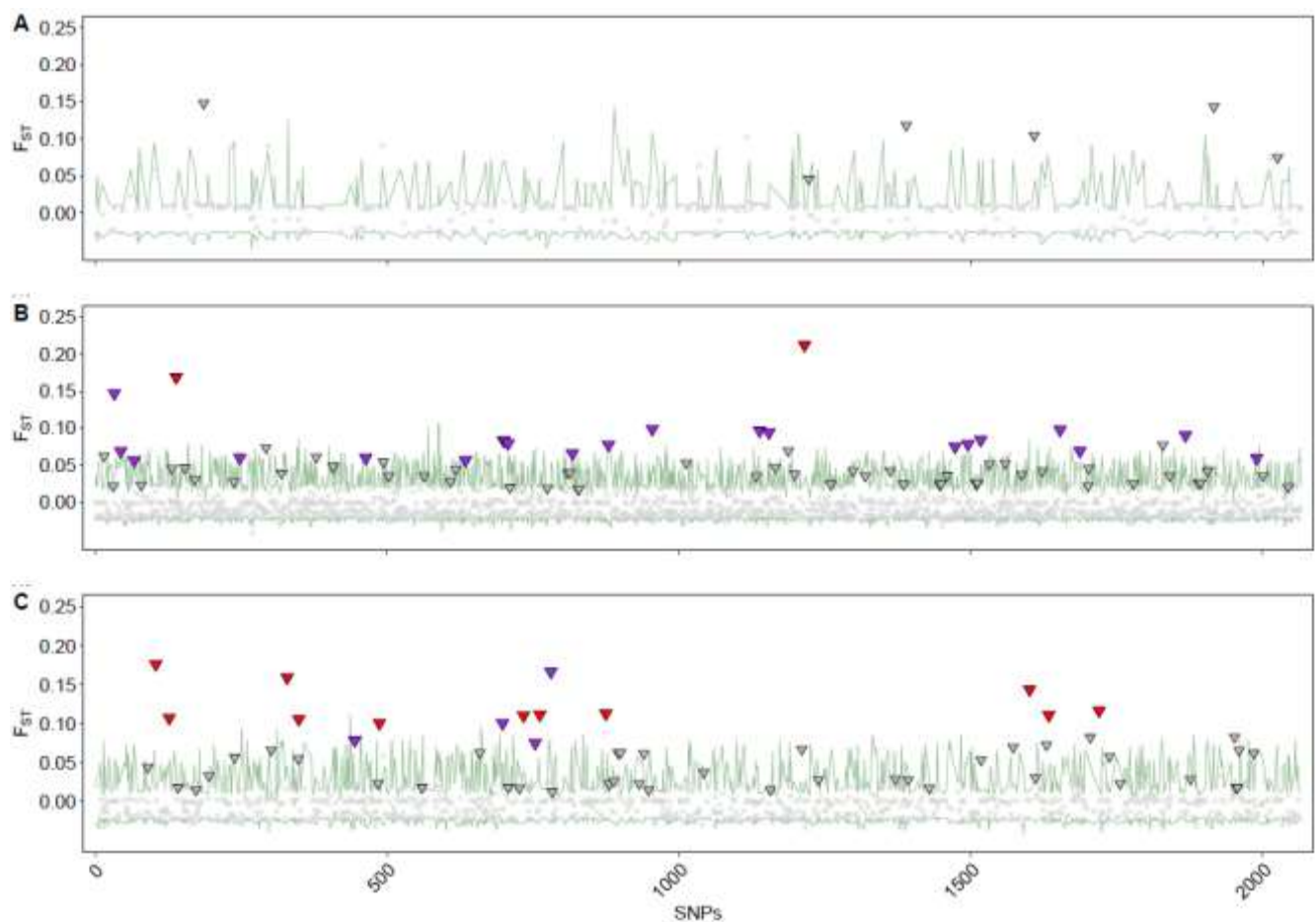

**Figure 5.** Outlier SNPs based on temporal differentiation are shown, depicting (A) the unique or shared outlier SNPs among methods for HLG, ROS, and QUI. The (B) allele frequency between time points and (C) observed $F_{ST}$ values are provided depending on whether the SNPs were detected by none or only one method (gray), two methods (2X; purple) and three methods (3X; red). Note that the total number of SNPs on the Venn diagrams (A) is not equal to 149 SNPs, as two outliers were detected twice in ROS and QUI by the simulation framework.

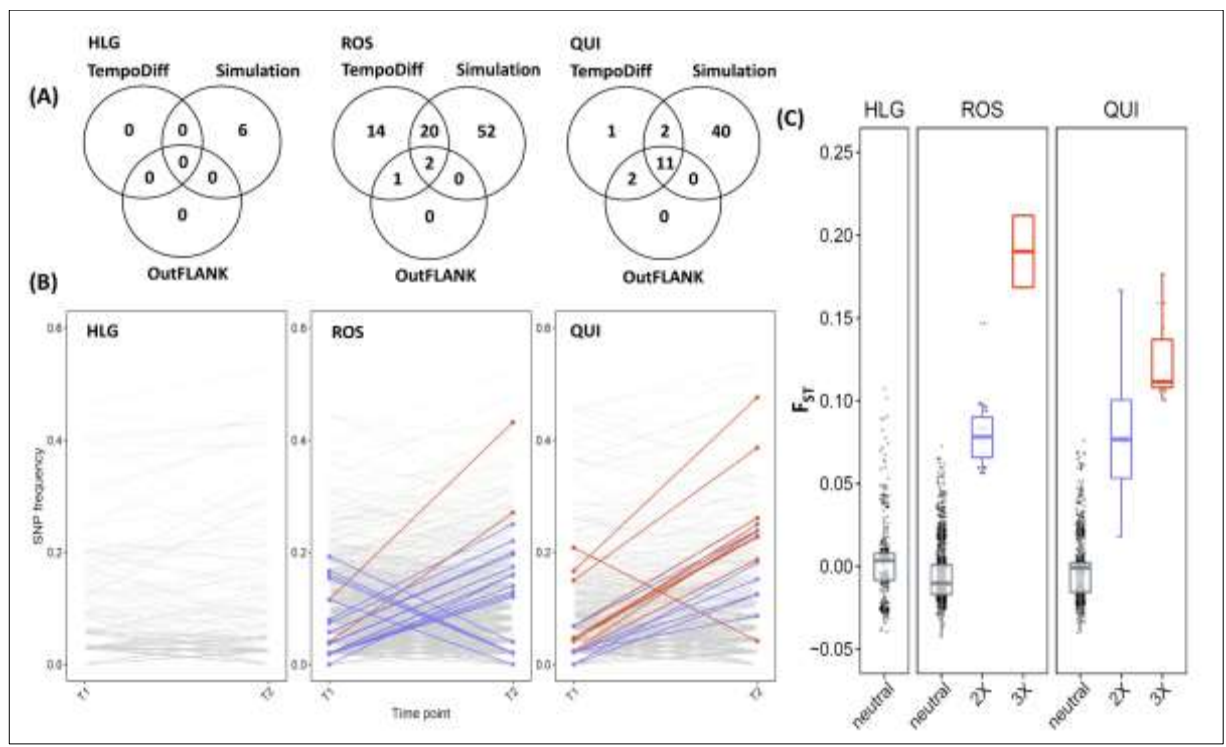

**Figure 6.** Absence (coded 0) or presence (coded 1) of the alternative allele for three outlier SNPs of the spatial analysis, best fitted by the highest and lowest mean SST predictors in the GLM models. Dots, colored according to the legend, represent observed data, while the response curve derived from the model highlights the relationship between SST and the alternative allele.

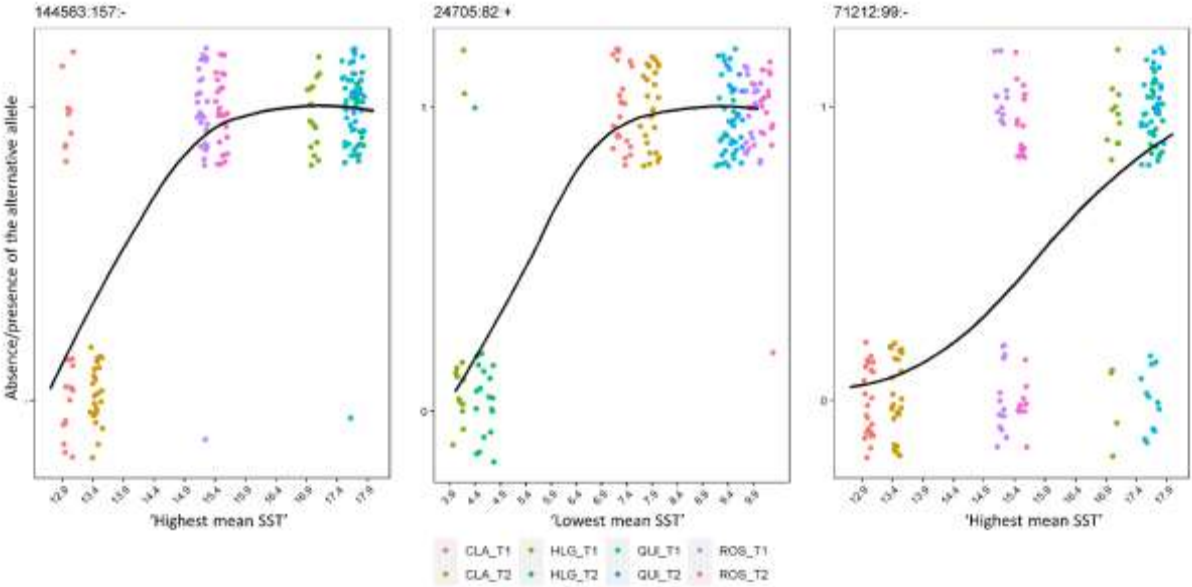

**Table 1.** Geographical coordinates and genetic diversity of populations at two different time points.

| Population | Year | Lat. | Lon. | n | P (± SE) | He (± SE) |
|---|---|---|---|---|---|---|
| Helgoland (HLG) | *2005* | 54.178 N | 7.893 E | 18 | 0.11 (0.001) | 0.015 (0.001) |
| | *2018* | | | 22 | 0.08 (0.001) | 0.013 (0.001) |
| Clachan (CLA_1) | *2008* | 56.454 N | 5.444 W | 25 | 0.42 (0.001) | 0.103 (0.003) |
| Clachan (CLA_2) | *2018* | 56.318 N | 5.592 W | 25 | 0.42 (0.001) | 0.109 (0.003) |
| Roscoff (ROS) | *2011* | 48.727 N | 4.005 W | 27 | 0.36 (0.001) | 0.065 (0.002) |
| | *2018* | | | 25 | 0.36 (0.001) | 0.069 (0.002) |
| Quiberon (QUI) | *2009* | 47.470 N | 3.091 W | 24 | 0.21 (0.001) | 0.038 (0.001) |
| | *2018* | | | 24 | 0.20 (0.001) | 0.040 (0.002) |

Year: the specific year when the samples were collected; Lat: latitude coordinates of the sampling location; Lon: longitude coordinates of the sampling location; n: the number of genotyped individuals; P (± SE): the proportion of polymorphic loci using a minor allele frequency threshold of 0.01 (mean and ± SE); He (± SE): expected heterozygosity, represented as the mean value with standard error (SE)

**Table 2.** The filtering steps applied to the raw sets of 77 193 SNPs and 644 794 SNPs for the first and second time points, respectively.

| Filtering steps | SNP | Individual |
|---|---|---|
| SNPs present in both the first and second time points | 36 293 | 213 |
| Shared SNPs among populations with less than 10% of missingness per SNP within a population | 4 151 | - |
| Individual missingness rate ≤ 0.2 | 4 151 | 190 |
| Mean read depth (DP); SNPs between 10x and twice the mean. For the first time point, the mean read depth was 34x, and for the second time point, it was 45x | 3 995 | - |
| SNPs deviating from HWE (P < 0.001) in at least 25% of the populations | 3 995 | - |
| Minimum allele count of three | 3 609 | - |
| SNPs pruned at an LD threshold of $r^2$ = 0.20 within and among populations | 2 854 | - |

**Table 3.** SST predictors employed in the analysis of SNP-temperature associations. SST predictors were estimated for each period according to the Pruned Exact Linear Time (PELT) algorithm.

| Population | Period | mean SST | highest mean SST | lowest mean SST |
|---|---|---|---|---|
| CLA | 1981-1996 | 10.07 | 12.99 | 7.32 |
| CLA | 1997-2018 | 10.65 | 13.46 | 7.87 |
| HLG | 1981-2004 | 10.34 | 17.02 | 3.99 |
| HLG | 2005-2018 | 11.08 | 17.74 | 4.58 |
| ROS | 1981-1995 | 12.37 | 15.18 | 9.92 |
| ROS | 1996-2018 | 12.85 | 15.51 | 10.27 |
| QUI | 1981-1993 | 13.44 | 17.57 | 9.35 |
| QUI | 1994-2018 | 13.68 | 17.75 | 9.51 |

*mean SST: monthly average over the entire period; highest mean SST: monthly average over the warmest months of the year (June, July, and August); lowest mean SST: monthly average over the coolest months of the year (January, February and March).

**Table 4.** Analyses of molecular variance (AMOVA) using the complete SNPs dataset

| Source of variation | D.f | Var. components | Variation (%) | P-value |
|---|---|---|---|---|
| Among populations | 4 | 16.07 | 20.71 | 0.00188 |
| Between time points | 3 | -0.47 | -0.61 | 0.70832 |
| Within populations | 182 | -0.88 | -1.14 | 0.91881 |
| Within individuals | 190 | 62.93 | 81.04 | < 0.00001 |

**Table 5.** Contemporary effective population size (Ne) based on temporal genetic changes using two F-statistics ($F_c$, $F_s$) and the likelihood estimator implemented in MLNe. Ne estimates were computed for the entire dataset and selectively for putatively neutral SNPs, achieved by excluding those identified through outlier tests in space.

| SNP dataset | Ne estimator | HLG | ROS | QUI |
|---|---|---|---|---|
| **initial dataset (2854 SNPs)** | *Fc* | 59 [26 - 469] | 451 [157 - Inf] | 165 [77 - 3412] |
| | *Fs* | 45 [23 - 1403] | 247 [111 - Inf] | 104 [56 - 738] |
| | *likelihood* | 104 [70 - 195] | 43 063 [935 - Inf] | 227 [134 - 667] |
| **Neutral (2838 SNPs)** | *Fc* | 77 [30 - Inf] | 443 [155 - Inf] | 168 [77 - 6558] |
| | *Fs* | 60 [28 - Inf] | 251 [112 - Inf] | 107 [57 - 918] |
| | *likelihood* | 112 [74 - 227] | 43 455 [949 - Inf] | 238 [138 - 772] |

# Supplementary section

## Supplementary material S1

This supplementary section aims to evaluate the impact of choosing low or high Ne scenarios in the simulation framework. We performed simulations of SNP frequencies 5 000 times with the same parameters as the ones described in the "Outlier detection across time" section of the main manuscript but considering the lowest Ne estimates obtained from temporal methods (see Table 5). We expected that decreasing Ne would strengthen genetic drift, leading to increased levels of temporal neutral genetic differentiation and a higher detection threshold for outlier SNPs. However, we found no significant differences in mean $F_{ST}$ values when comparing simulations based on the low and high Ne scenarios. The mean $F_{ST}$ values ranged from 0.002 to 0.003 for the high and low Ne scenarios, respectively. Additionally, the detection of outlier SNPs was minimally influenced by this parameter. Person's correlations between SNP p-values in the low and high Ne scenarios were high, with correlations of r = 0.962 at HLG, r = 0.997 at ROS and r = 0.986 at QUI. This indicates that the SNPs detected as outliers were almost identical across the two different Ne scenarios. Interestingly, reducing Ne at ROS from 43 063 to 247 prevented the detection of two outliers that were reported from the high Ne scenario. It is noteworthy that major differences were expected at ROS, considering the reduction in Ne from thousands to hundreds of individuals, while the range was narrower at HLG and QUI (around two times). This suggests that changes in allele frequencies under genetic drift over two generations may be buffered at ROS by maintaining Ne in at least two hundred individuals. However, subsequent analyses indicated that reducing Ne had a greater impact on simulations and genetic changes when genetic drift was simulated over a thousand generations, rather than two (data not shown).

**Figure S1.** SNMF admixture plots for different values of K, including K = 2, K = 3, and the optimal number of clusters K = 5. Each bar in the plots represents one sampled individual, and the colors within the bar represent the membership proportions of each cluster.

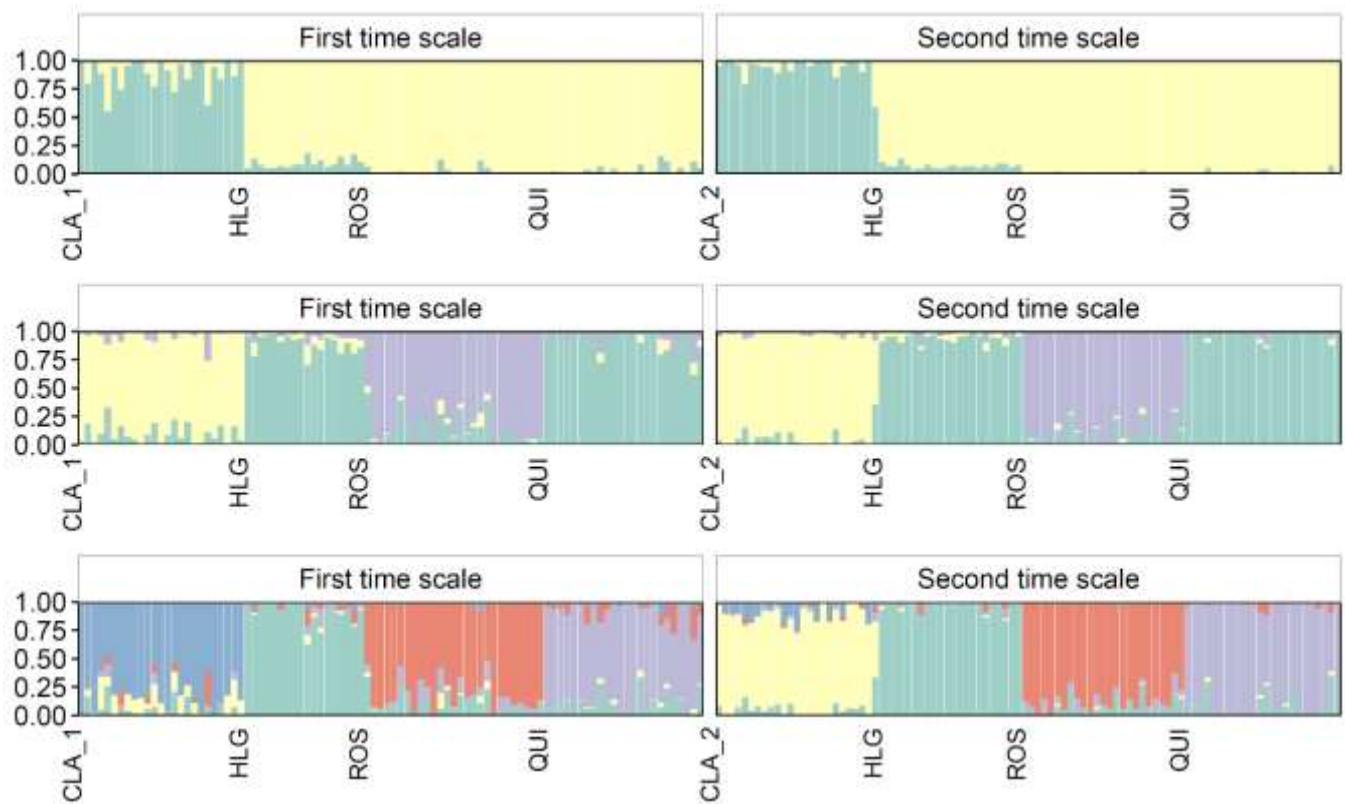

**Figure S2.** PCA biplots of SNP variation, including the five principal components: PC1 (46.5%) vs PC2 (8.3%); PC2 vs PC3 (6.1%), PC3 vs PC4 (4.5%) and PC4 vs PC5 (2.7%).

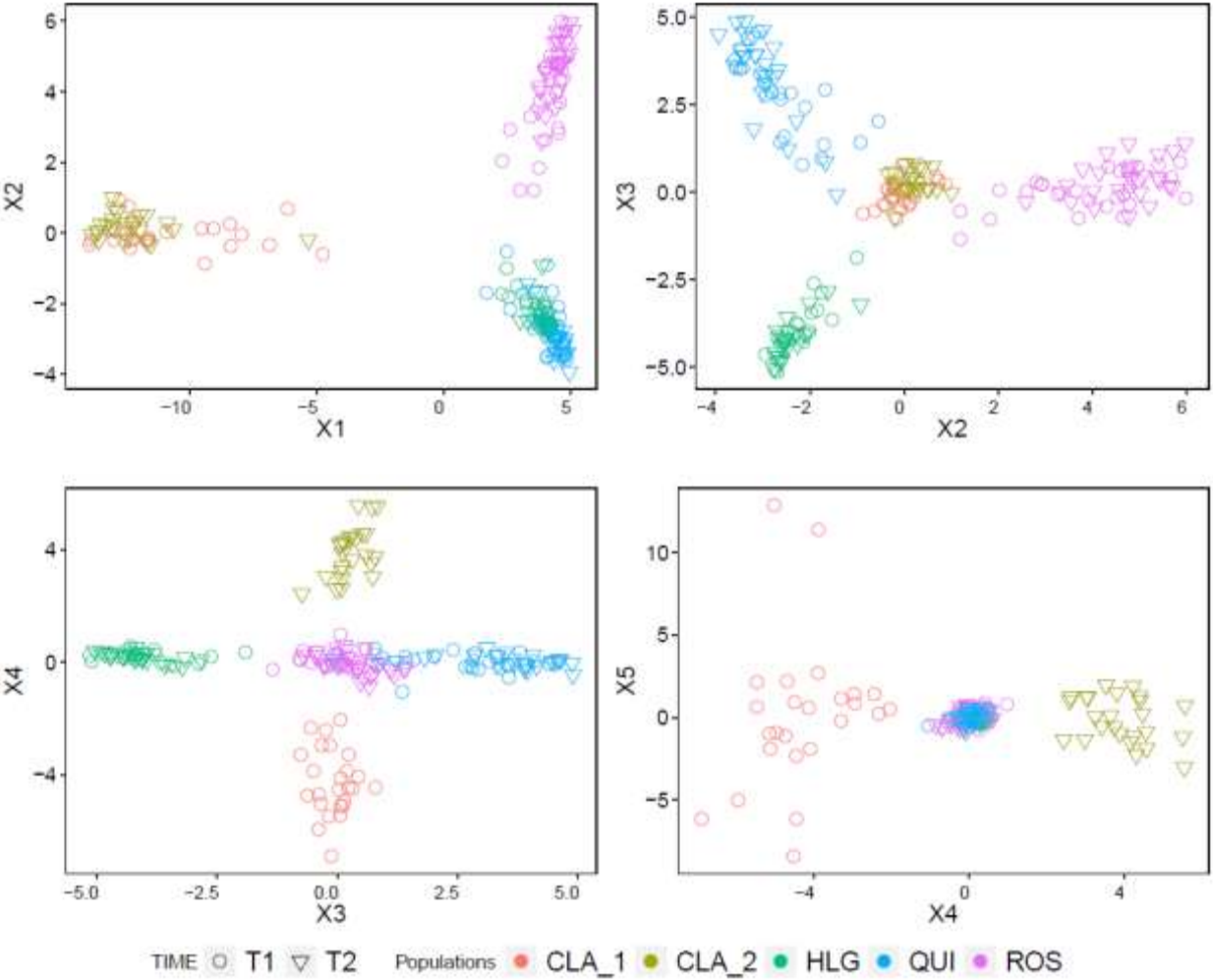

**Figure S3.** Analysis of minor allele frequencies (MAF) for temporal outliers reported at Quiberon (QUI) and Roscoff (Roscoff). The upper plots illustrate the density of the MAF at the first time point, while the lower plots report those at the second time point for the same SNPs.

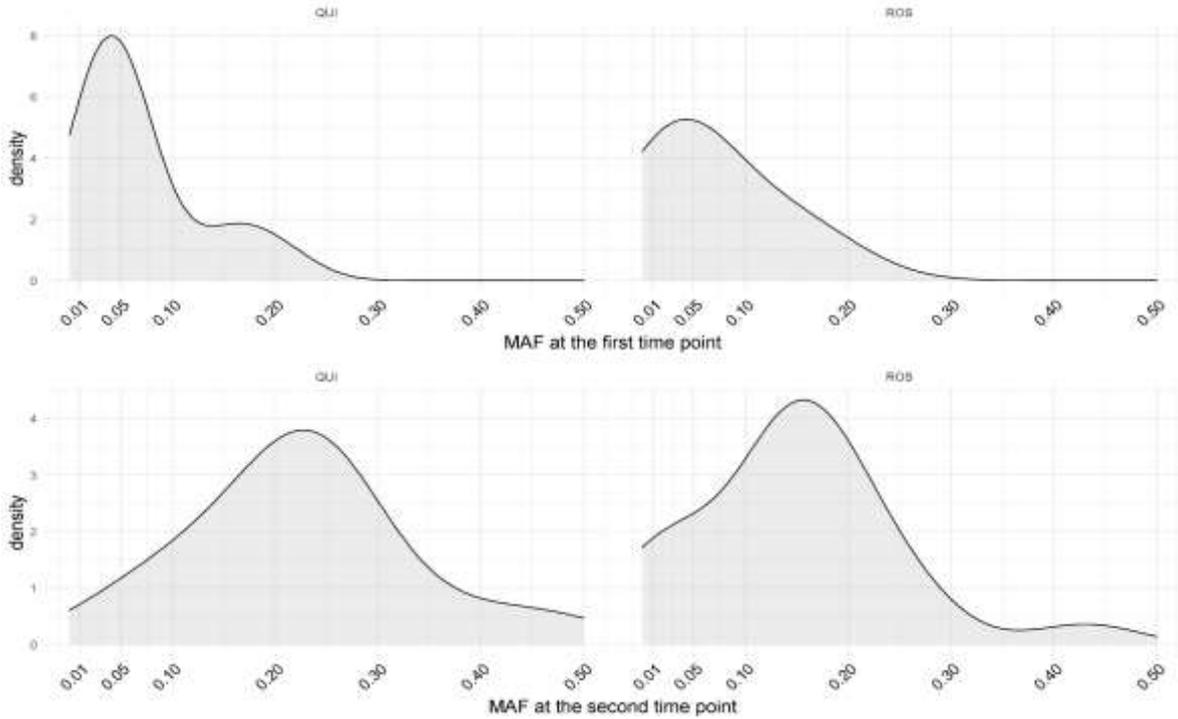

**Table S1**. Individual missingness, and mean read depth (Mean DP) were assessed at both time points. Individual missingness refers to the average proportion of missing genotypes for each individual. Mean read depth was calculated by averaging the read depth across all sites. The SNP error rate estimated using eight replicates is also indicated.

| Population | Time point | n | Ind. missingness | Mean DP | SNP error rate |
|---|---|---|---|---|---|
| HLG | 1st | 18 | 0.05 | 13.218 | 0.020 |
| | 2nd | 22 | 0.02 | 11.640 | 0.012 |
| CLA | 1st | 25 | 0.04 | 12.497 | 0.036 |
| | 2nd | 25 | 0.02 | 26.362 | 0.002 |
| ROS | 1st | 27 | 0.03 | 28.457 | 0.021 |
| | 2nd | 25 | 0.02 | 26.177 | 0.001 |
| QUI | 1st | 24 | 0.06 | 10.570 | 0.007 |
| | 2nd | 24 | 0.02 | 24.703 | 0 |

**Table S2**. Pairwise $F_{ST}$ values between populations and the comparisons between time points are highlighted in bold in the lower diagonal matrix. The p-values of the genic differentiation tests are displayed in the upper diagonal matrix.

| $F_{ST}$ | HLG (T1) | HLG (T2) | CLA_1 | CLA_2 | ROS (T1) | ROS (T2) | QUI (T1) | QUI (T2) |
|---|---|---|---|---|---|---|---|---|
| HLG (T1) | | **NS** | < 0.001 | < 0.001 | < 0.001 | < 0.001 | < 0.001 | < 0.001 |
| HLG (T2) | **0.008** | | < 0.001 | < 0.001 | < 0.001 | < 0.001 | < 0.001 | < 0.001 |
| CLA_1 | 0.257 | 0.285 | | **< 0.001** | < 0.001 | < 0.001 | < 0.001 | < 0.001 |
| CLA_2 | 0.266 | 0.297 | **0.039** | | < 0.001 | < 0.001 | < 0.001 | < 0.001 |
| ROS (T1) | 0.094 | 0.113 | 0.224 | 0.235 | | **NS** | < 0.001 | < 0.001 |
| ROS (T2) | 0.104 | 0.125 | 0.223 | 0.237 | **0.000** | | < 0.001 | < 0.001 |
| QUI (T1) | 0.127 | 0.150 | 0.251 | 0.262 | 0.078 | 0.083 | | **NS** |
| QUI (T2) | 0.142 | 0.167 | 0.249 | 0.272 | 0.090 | 0.095 | **0.003** | |

**Table S3.** Predictors retained in each GLM model for predicting the occurrence of the alternative allele in outlier SNPs detected in space. Only predictors deemed significant after conducting a stepwise variable selection procedure are presented. P-values highlighted in bold indicate predictors that remain significant after correcting for multiple testing.

| SNP ID | fitted model | OR | IC2.5 | IC97.5 | Pval | AIC | R2 |
|---|---|---|---|---|---|---|---|
| 144563:157:- | PC4 | 0.54 | 0.29 | 0.74 | 0.0053 | 58.07 | 0.74 |
| 144563:157:- | highest mean SST | 8.03 | 4.42 | 21.7 | **<0.003** | 58.07 | 0.74 |
| 174791:33:+ | PC1 | 2.29 | 1.62 | 5.92 | 0.0045 | 46.02 | 0.83 |
| 218088:45:+ | PC1 | 1.65 | 1.45 | 2.01 | **<0.003** | 36.73 | 0.84 |
| 24705:82:+ | lowest mean SST | 7.68 | 3.3 | 104.45 | 0.0043 | 39.54 | 0.82 |
| 35776:116:+ | PC1 | 1.63 | 1.4 | 2.21 | **<0.003** | 70.05 | 0.72 |
| 59182:127:- | PC1 | 1.93 | 1.57 | 2.94 | **<0.003** | 24.45 | 0.91 |
| 71212:99:- | PC1 | 11.06 | 4.13 | 37.49 | **<0.003** | 118.57 | 0.56 |
| 71212:99:- | PC3 | 0.67 | 0.48 | 0.87 | 0.0077 | 118.57 | 0.56 |
| 71212:99:- | PC5 | 0.11 | 0.02 | 0.45 | 0.0033 | 118.57 | 0.56 |
| 71212:99:- | highest mean SST | 3.16 | 1.94 | 5.67 | **<0.003** | 118.57 | 0.56 |
| 72447:52:- | PC2 | 0.58 | 0.32 | 0.78 | 0.0050 | 58.14 | 0.77 |

[*] This summary includes Odds Ratios (OR), estimated odds ratios derived from logistic regression coefficients, 95% Confidence Intervals (CIs) for ORs, P-values, AIC, and pseudo $R^2$ values of the fitted model.

1    **Table S4.** Functional categorization of eight annotated loci derived from outlier SNPs out of the 53 loci.

| Locus | Description | Length | Hits | e-value | sim mean | GO | GO Names | Enzyme Names | 3 |
|-------|-------------|--------|------|---------|----------|-----|----------|--------------|---|
| 71212 | conserved unknown protein | 216 | 2 | 6.4E-19 | 92.0 | 2 | P:lipid transport; C:cellular anatomical entity | | 4 |
| 100680 | ankyrin repeat protein | 254 | 20 | 4.8E-11 | 54.24 | 3 | P:protein phosphorylation; F:protein kinase activity; F:ATP binding | Transferring phosphorus-containing groups | 5 |
| 10940 | MutS protein homolog 4 | 503 | 1 | 1.9E-4 | 85.29 | 4 | P:mismatch repair; F:ATP binding; F:mismatched DNA binding; F:ATP-dependent DNA damage sensor activity | | 6 |
| 30235 | ankyrin repeat protein | 241 | 20 | 8.1E-10 | 54.25 | 1 | C:membrane | | |
| 59182 | 2-hydroxy-3-oxopropionate reductase | 339 | 2 | 6.9E-4 | 96.3 | 3 | F:oxidoreductase activity; F:NADP binding; F:NAD binding | Oxidoreductases | |
| 62443 | unnamed protein product | 319 | 5 | 1.6E-6 | 60.99 | 5 | P:RNA polyadenylation; F:polynucleotide adenylyltransferase activity; F:metal ion binding; C:membrane; C:TRAMP complex | polynucleotide adenylyltransferase | |
| 94321 | unnamed protein product | 229 | 1 | 2.8E-4 | 95.45 | | | | |
| 95165 | methyltransferase | 257 | 19 | 2.2E-5 | 71.64 | | | | |